\documentclass[amsmath,amssymb]{webofc}

\usepackage[varg]{txfonts}   
\usepackage{hyperref}
\usepackage{url}
\usepackage{bm}
\hypersetup{colorlinks=true,citecolor=blue,urlcolor=blue,linkcolor=blue}
\begin{document}
\title{Hot and dense pQCD in a very strong magnetic background }
%
%

\author{\firstname{Eduardo S.}
\lastname{Fraga}\inst{1}\fnsep\thanks{\email{fraga@if.ufrj.br}} \and
        \firstname{Let\'icia F.}
\lastname{Palhares}\inst{2} \and
        \firstname{Tulio E.}
\lastname{Restrepo}\inst{3}
}

\institute{ Instituto de F\'isica, Universidade Federal do Rio de Janeiro,
Caixa Postal 68528, 21941-972, Rio de Janeiro, RJ, Brazil
\and
           Universidade do Estado do Rio de Janeiro, Instituto de F\'isica,
Departamento de F\'isica Te\'orica,
Rua S\~ao Francisco Xavier 524, 20550-013 Maracan\~a, Rio de Janeiro, Brazil
\and
           Department of Physics, University of Houston, Houston, TX 77204, USA
          }

\abstract{We compute the pressure, chiral condensate and strange quark number
susceptibility from perturbative QCD up to two-loop order at finite
temperature and very high magnetic fields  with physical quark
masses. We also discuss the case of cold and dense quark matter in the presence of a very strong magnetic field and constraints for quark magnetars. }
\maketitle
\section{Introduction}
\label{intro}
Magnetic quantum chromodynamics (QCD) is relevant in different scenarios, from magnetars \cite{Kaspi:2017fwg} to non-central heavy-ion collisions \cite{Kharzeev:2007jp} and the early universe \cite{Vachaspati:1991nm}. Lattice QCD  calculations can be performed at zero density, where there is no sign problem \cite{Bali:2013esa}. 
For large enough values of the running scale one can also rely on perturbation theory. In Ref. \cite{Blaizot:2012sd} the two-loop expression of the pressure was calculated in the lowest Landau level (LLL) approximation, and its dependence on the current quark mass was studied for fixed values of temperature, scale, and coupling. Ten years later, we revisited this problem in the case of zero density \cite{Fraga:2023cef}, providing a more compact and numerically convenient expression, and using a running scale, $\Lambda=\sqrt{(2\pi T)^2+eB}$, for the coupling and strange quark mass. We computed the pressure, quark number susceptibility, and renormalized quark condensate at very high magnetic fields comparing them to recent lattice results \cite{DElia:2021yvk}. We also extended the results to the case of cold and dense quark matter, computing properties of quark magnetars \cite{Fraga:2023lzn}. Recently, we calculated Taylor coefficients of the pressure \cite{Fraga:2025juh} and compared them to lattice data \cite{Astrakhantsev:2024mat}.

\section{Hot pQCD}
\label{sec-1}
At very high magnetic fields, where $m_s \ll T \ll \sqrt{eB}$, one finds that only the LLL contributes. Using this approximation, the free (one-loop) pressure reads
\begin{align}
\begin{split}
 \frac{P_{\rm free}^{\rm LLL}}{N_c}=&
 -\sum_f\frac{(q_fB)^2}{2\pi^2}\left[x_f\ln\sqrt{x_f}\right]+T\sum_{f}\frac{q_fB}{2\pi}\int \frac{dp_z}{2\pi}\bigg \{\ln\left(1+e^{-\beta\left[E(p_z)-\mu_f\right]}\right)+\ln\left(1+e^{-\beta\left[E(p_z)+\mu_f\right]}\right)\bigg \} \, ,
 \end{split} \label{Pfree}
\end{align}
where $E^2=p_z^2+m^2_f$, $x\equiv m_f^2/(2q_fB)$, $q_f$ is the electric charge of the quark of flavor $f$, $m_f$ its respective current mass, $T=1/\beta$ is the temperature, and $\mu_f$ is the quark chemical potential.

\begin{figure}[h]
\centering
\includegraphics[width=5cm,clip]{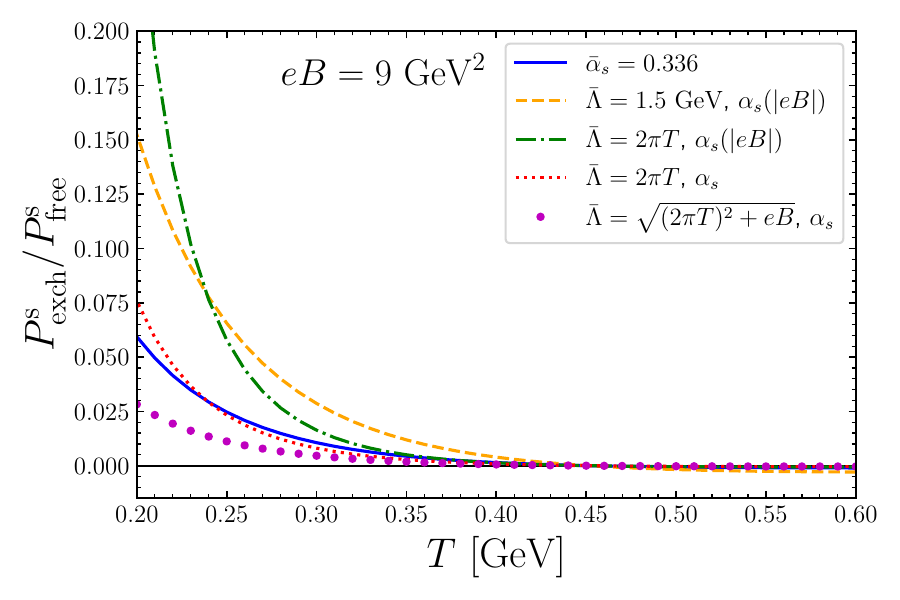}
\caption{$P^s_{\rm exch}/P^s_{\rm free}$ as functions of the temperature for $eB=9$ $\rm{GeV}^2$.}
\label{fig-1}       
\end{figure}

The thermal exchange pressure can be conveniently written as \cite{Fraga:2023cef}
\begin{align}
\begin{split}
\frac{P_{\rm exch}^{\rm LLL}}{N_c}=&\frac{1}{2}g^2 \left(\frac{N_c^2-1}{2N_c}\right)T^2 \sum_f m_f^2\left(\frac{q_fB}{2\pi}\right)\sum_{\ell,n_2}\int \frac{dm_k}{2\pi}m_k
 e^{-\frac{m_k^2}{2 q_f B}}\frac{\mathcal{E}_\ell-\mathcal{E}_{n_2}}{\mathcal{E}_\ell \mathcal{E}_{n_1} \mathcal{E}_{n_2} \left|\mathcal{E}_\ell-\mathcal{E}_{n_2}\right| \left(\left|\mathcal{E}_\ell-\mathcal{E}_{n_2}\right|+\mathcal{E}_{n_1}\right)} \, ,
\end{split}\label{Pexch}
\end{align}
where $\mathcal{E}_\ell=\sqrt{\omega_\ell^2+m_k^2}$, $\mathcal{E}_{n_1}=\sqrt{(\omega_{n_2}+\omega_\ell)^2+m_f^2}$, $\mathcal{E}_{n_2}=\sqrt{\omega_{n_2}^2+m_f^2}$, $\omega_\ell = 2\pi \ell T$ and $\omega_{n_2}=(2n_2+1)\pi T$.
In Figure \ref{fig-1}, we plot the ratio $P^s_{\rm{exch}}/P^s_{\rm{free}}$ of the strange pressure as a function of temperature for $eB=9$ $\rm{GeV}^2$ and for different expressions of the running coupling, namely: $\alpha_s=0.0336$; the $B$-dependent expression of Ref. \cite{Ayala:2018wux} with a constant scale $\Lambda=1.5$ GeV; the same expression with $\Lambda=2\pi T$; usual 2L expression for the coupling with $\Lambda=2\pi T$; and two-loop coupling with $\Lambda=\sqrt{(2\pi T)^2+eB}$. 
The quark mass $m_s(\Lambda)$ runs with the usual two-loop expression.
We can see in figure \ref{fig-1} that the exchange contribution to the pressure is negligible at high $T$, and with the exception of the case with the coupling from Ref \cite{Ayala:2018wux}, the exchange pressure contributes very little at low $T$, suggesting a better convergence than pQCD at $eB=0$.
This results constrain the behavior at high density and very large magnetic field. 
\begin{figure}[h]
\centering
    \includegraphics[width=5cm,clip]{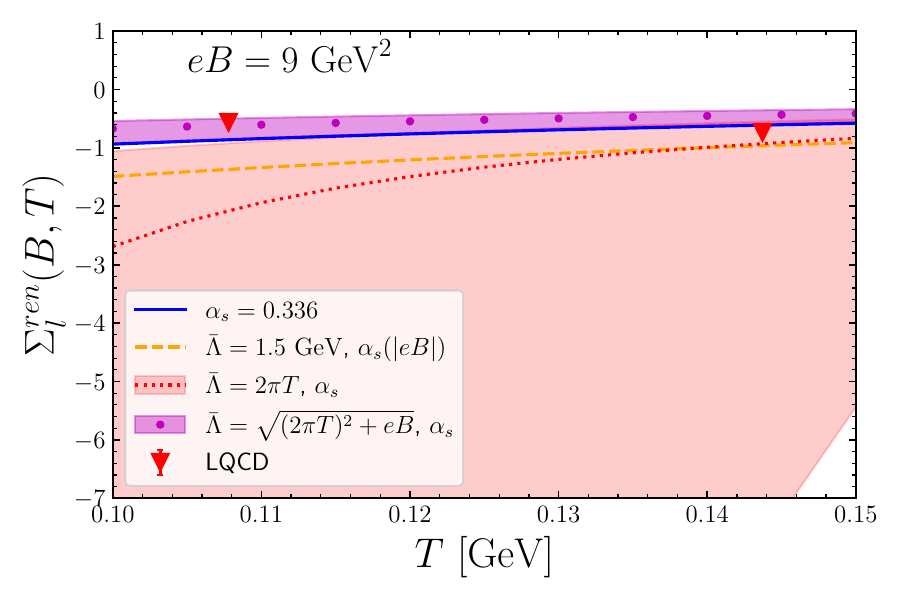}
    \includegraphics[width=5cm,clip]{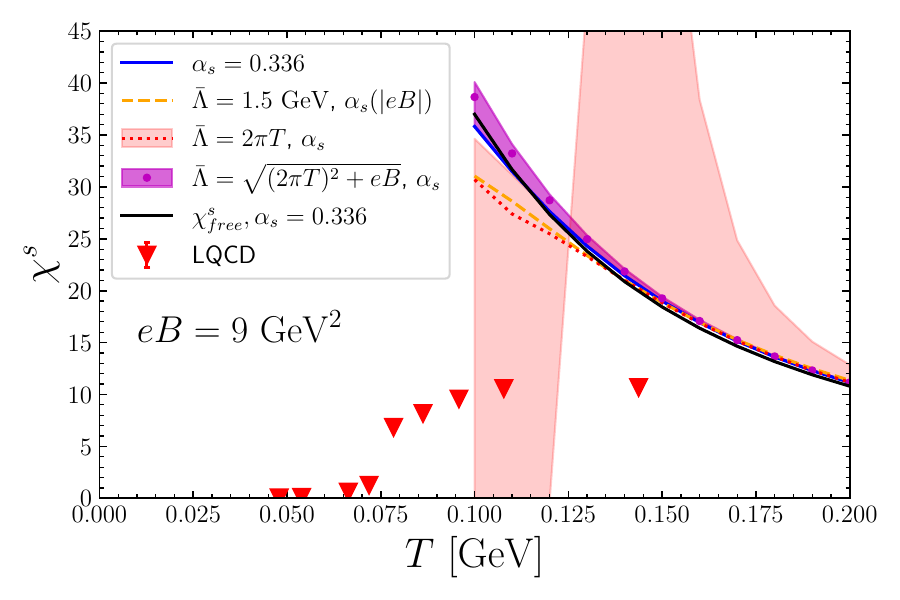}
\caption{Normalized light quark condensate (left) and strange quark number susceptibility (right). 
The bands correspond to changes in the central scale by a factor of 2. Lattice data from Ref. \cite{DElia:2021yvk}}
\label{fig-2}       
\end{figure}
In figure \ref{fig-2}, we show the normalized light quark condensate (left) and the strange quark number susceptibility (right) for $eB=9\,\rm {GeV}^2$. 
Our results are compared with lattice data from Ref. \cite{DElia:2021yvk}. 
Even though the temperature range is rather limited, our results are in the same ballpark as the lattice at higher values of temperature. 
\section{Cold and dense pQCD}
\label{sec-2}
Taking the limit $T \to 0$ in cold and dense perturbative QCD at high magnetic fields simplifies the expressions from Ref. \cite{Blaizot:2012sd}, yielding the result \cite{Fraga:2023lzn}.
\begin{align}
 \frac{P_{\rm free}^{\rm LLL}}{N_c}= -\sum_f\frac{(q_fB)^2}{2\pi^2}\left[x_f\ln\sqrt{x_f}\right]+\sum_{f}\frac{(q_fB)}{4\pi^2}\left[ \mu_f P_F - m_f^2 \log\left( \frac{\mu_f+P_F}{m_f} \right) \right] 
 \, ,
  \label{Pfree_T0}
\end{align}
and 
\begin{align}
\begin{split}
 \frac{P_{\rm exch}^{\rm LLL}}{N_c}=&-\frac{1}{2}g^2 \left(\frac{N_c^2-1}{2N_c}\right) m_f^2\left(\frac{q_fB}{2\pi}\right)\int \frac{dm_k}{2\pi}m_ke^{-\frac{m_k^2}{2 q_f B}}\int \frac{dp_zdq_zdk_z}{(2\pi)^3}(2\pi)\delta(k_z-p_z+q_z)\\
 &\times\frac{1}{\omega E_p E_q}\Bigg\{\frac{\omega}{E_-^2-\omega^2}\Theta(\mu_f-E_{\bf{p}})\Theta(\mu_f-E_{\bf{q}}) -\left[\frac{2\left(E_{\bf{q}}+\omega\right)}{\left(E_--\omega\right)\left(E_++\omega\right)}\right]\Theta(\mu_f-E_{\bf{p}})-\frac{1}{E_{+}+\omega}\Bigg\} \, .
\end{split}\label{P_exch_T_finite}
\end{align}
\begin{figure}[h]
\centering
\includegraphics[width=5cm,clip]{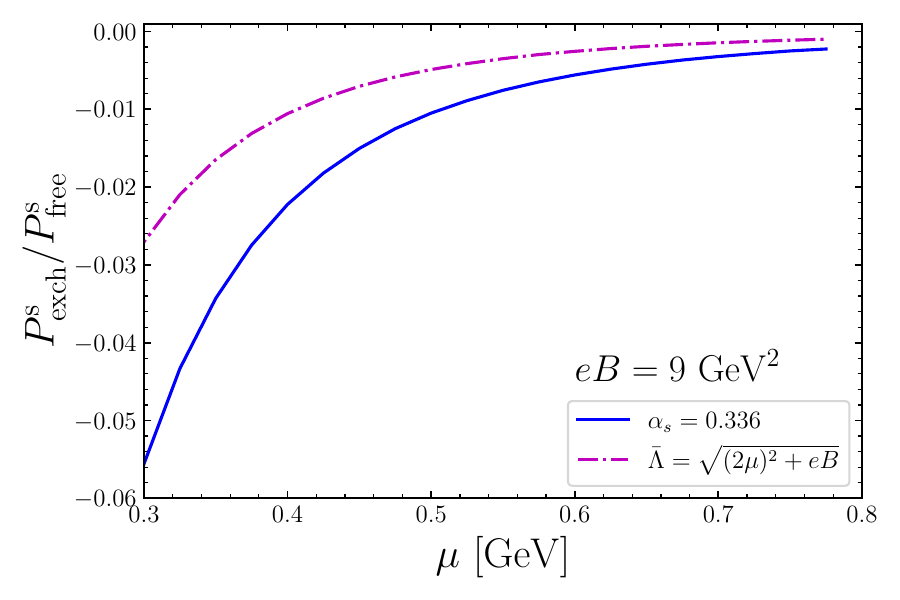}
\caption{$P^s_{\rm exch}/P^s_{\rm free}$ as a function of the chemical potential for $eB=9$ $\rm{GeV}^2$.}
\label{fig-3}       
\end{figure}
As in the previous case, the exchange contribution is negligible compared with the free pressure, as shown in Fig. \ref{fig-3}. This allows for the construction of a simple analytic model in which the pressure is given by the ``free'' pressure with running mass \cite{Fraga:2023lzn}.
\subsection{Quark magnetars}
We can now compute the mass-radius diagram of quark magnetars, with the total pressure of the system given by 
$P=P_{\rm eff}+W-B^2/2+P_e$. 
Here, $P_{\rm eff}$ denotes the free quark pressure Eq. (\ref{Pfree_T0}), $-B^2/2$ represents the purely magnetic contribution, $P_e$ is the free electron pressure, and $W$ is a function that ensures thermodynamic consistency \cite{Restrepo:2022wqn}.

\begin{figure}[h]
\centering
    \includegraphics[width=5cm,clip]{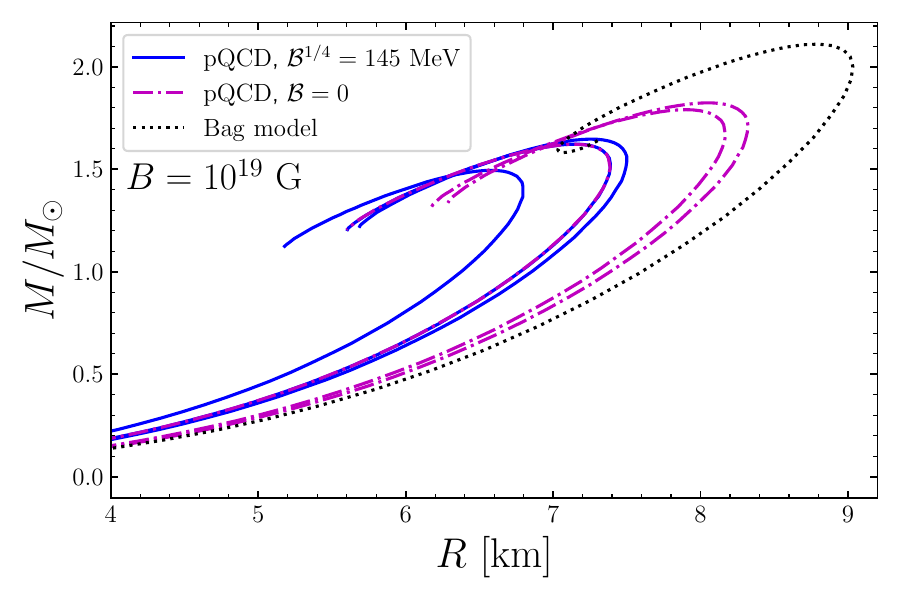}
    \includegraphics[width=5cm,clip]{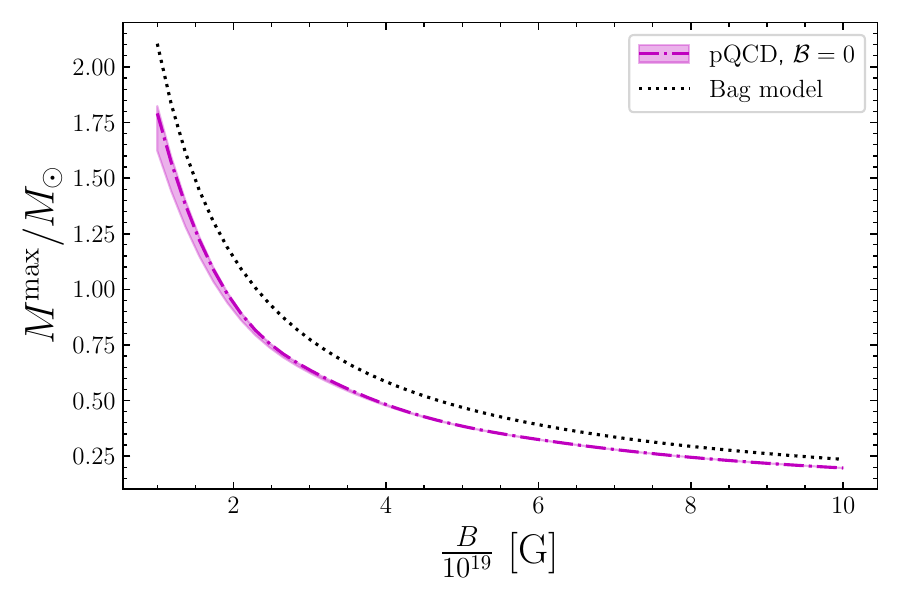}
\caption{Mass-radius of quark magnetars (left) and their total mass as a function of $B$ (right).}
\label{fig-4}       
\end{figure}
Figure \ref{fig-4} shows the mass-radius relation (left) and the maximum mass (right) of quark magnetars. We show results from pQCD with and without a bag constant, $\mathcal{B}$, and compare them with a simple bag model, $P_{\rm bag} = \epsilon - 2\mathcal{B} + B$, where $\epsilon$ is the energy density. Our results predict slightly smaller maximum masses than those of the bag model. Increasing $B$ steeply reduces the maximum mass values.
\section*{\acknowledgementname}{This work was partially supported by CAPES (Finance Code 001), CNPq, FAPERJ, and INCT-FNA (Process No. 464898/2014-5).}

%
%
%
\bibliography{refs.bib}





\end{document}